\documentclass{aastex}
\usepackage{emulateapj5}
\newcommand\simlt{\lower.5ex\hbox{$\; \buildrel < \over \sim \;$}}
\newcommand\simgt{\lower.5ex\hbox{$\; \buildrel > \over \sim \;$}}

\begin{document}
%\large
%\begin{titlepage}
\title{Compton Sailing and Strong Polarization}
\author{ David Eichler\altaffilmark{1}}
\altaffiltext{1}{Physics Department, Ben-Gurion University,
Beer-Sheva 84105, Israel; eichler@bgumail.bgu.ac.il}

\begin{abstract}
It is noted that a surface layer of matter in contact with a
sufficiently super-Eddington, radially combed photon flux
typically  attains a relativistic coasting state whereby the
radiation does not accelerate the matter. The final bulk Lorentz
factor of this matter is therefore determined by geometry.
Radiation that scatters off this layer is most likely to be
observed along the velocity vector of the matter, where it would
be most strongly polarized.
\end{abstract}

A remarkable feature of cosmologically distant gamma ray bursts is
their high isotropic equivalent luminosities, which can exceed the
Eddington limit by up to 14 orders of magnitude. Such a flux would
sharply accelerate baryonic matter that stood in its way and, even
for a small quantity of such matter,  most of the GRB energy would
be expended on its acceleration. This would require salvage
mechanisms such as internal shocks, which,  in order to be
effective in converting efficiently to gamma radiation, must occur
near or beyond the photosphere, i.e. at large radii.

Is it necessary to have any baryonic matter at all in GRB? If GRB
emerge along field lines that thread an event horizon, they need
not receive any baryons, and most of the pairs could annihilate at
relatively small radii when the temperature falls below
 $2 \times 10^8$K.  The indirect evidence
for matter in the fireballs  is a) non-thermal spectra, which
suggest that much of the energy passes through charged particles
at low optical depth, and b) the existence of afterglows, which
would not be generated purely by gamma rays. There are, however,
models that could yield non-thermal emission and afterglow without
a significant amount of baryons. For example, fireball energy is
easily converted to ultrahigh-energy particles if neutrons decay
within the baryon-poor jet after creeping in across field lines
(Eichler \& Levinson 1999, Levinson \& Eichler 2003). The
ultra-high energy particles could carry enough energy per electron
that they could deliver the required non-thermal gamma ray energy
and afterglow energy while remaining optically thin to the
non-thermal gamma rays. Other models  posit pair creation purely
from gamma rays (Thompson \& Madau 2000, Beloborodov 2001)
 or Poynting flux (Lyutikov, Parlev,  \& Blandford 2003) at large radii.
 The photosphere, which would be at large radius for a baryon
 dominated plasma, can in view of these non-baryonic alternatives be at much
 smaller radius. To quantify this, note that if the energy that remains kinetic - ultimately
destined to power the afterglow -  is carried as normal
electron-ion plasma, then the photosphere occurs at
$r=1/(1-\beta)\sigma_T n_e \sim
10^{12}L_{b50}(\frac{\Gamma}{10^2})^{-3}(\frac{\theta}{0.2})^{-2}$
cm. Here $L_{b50}$ is the kinetic luminosity of the baryons, $4\pi
\Gamma \gamma_p m_pn_ec^3 \theta^2 r^2$, in units of
$10^{50}$erg/s, the neutron to proton ratio in the ions is assumed
to be unity, and  we have tentatively included the possibility
that the internal Lorentz factor $\gamma_p$ for the ions may be
larger than unity. These considerations would together  suggest
that $r_{12} \ge 1$ and that $\Gamma \ge 10^2$, which is widely
though not universally believed  for GRB fireballs. If, on the
other hand, the fireball plasma is mostly pairs then the
photosphere can be dictated by pair recombination and  occur at
considerably smaller radius, ($r_{ph}\stackrel{\sim}{<}
10^{11}$cm), e.g. Eichler (1994), Eichler \& Levinson (2000).

A now familiar suggestion for getting kinetic fireball  energy
into gamma radiation, is via internal shocks near (Eichler 1994)
or downstream  of (Rees \& Mezsaros 1994) the gamma-ray
photosphere. The question still remains whether the soft gamma ray
emission, is emitted from optically thin regions or from a
photosphere, or both. The low frequency parts of GRB spectra are
frequently too steeply rising, it is reported (e.g. Preece et al.
2002), to be consistent with optically thin, efficient synchrotron
radiation. Alternative pictures, which might allow a steeper rise
in the soft part of the spectrum, involve Compton upscattering
(Shaviv \& Dar 1995, Lazzati et al. 2000) of softer photons by the
energy-bearing, relativistic outflow.  Here too, however, the bulk
and internal Lorentz factors must be fine tuned to get the right
peak energy.

However, there is another alternative: that radiation from or near
a photosphere is the primary energy source, this photosphere is
largely  determined by net annihilation of  positrons - which
which loosely associates  the peak energy with $m_ec^2$ - and
baryon involvement, while generating  additional effects,  occurs
too far downstream to spoil this association.  Recent evidence of
GRB collimation suggests that the fireball, though sufficiently
baryon-pure to allow high bulk Lorentz factors, is surrounded by
baryon-rich material either the envelope of a host star, or a wind
from the accretion disk (Levinson \& Eichler 2000). This allows an
entirely different set of models for GRB's, one in which the GRB
fireball would be pure photons (or Poynting flux) except for
matter that is "shaved off" the inner wall of the baryonic sheath
that collimates the fireball. The shavings could be accelerated to
relativistic Lorentz factors, and could play whatever role is
ascribed to matter in GRB fireballs, such as eventually supplying
non-thermal particles and generating afterglow.  However, they
need not be the primary source of soft gamma rays  and they need
not be subject to constraints on $\Gamma$ derived from time
variability.

The purpose of this Letter is to consider the interaction between
a radiative fireball and a sheath of baryonic matter when the
latter collimates the former. It is argued that the  moving wall
of matter that is in contact with the fireball  maintains a
Lorentz factor $\Gamma$ that is the inverse of the sine of the
impact angle between the photons and the outflow. This is due to
the fact that this matter "sails" on the radiation pressure. This
picture could yield an estimate of the Lorentz factor of the
matter that is associated with GRB fireballs. It also predicts
that the polarization is typically high; it can be close to 100
percent, and can exceed 60 percent with significant probability.
This is considerably higher on average than in  previous
suggestions (e.g. Shaviv \& Dar 1995,  Eichler \& Levinson 2003).

  Compton equilibrium between matter and radiation
pressure has previously been considered in the context of an
accretion disk by Sikora and Wilson (1981) Phinney (1982), Sikora et
al (1996). Begelman and Sikora (1987) considered polarization from
scattering off outflow, but did not specifically consider the
correlation of such polarization with Compton equilibrium.
Beloborodov (1998) considered the polarization vector in the case of
a 1-D outflow from a disk with Compton equilibrium at modest
$\beta$, $\beta\sim 0.5$. The polarization he found is along the
outflow axis, orthogonal  to what is expected in the context of the
present paper. The polarization effect proposed here is  a
consequence of high Lorentz factor combined with Compton
equilibrium.

Consider a  flux of $\frac{l L_{Edd}}{4\pi r^{2}}$, where $l\sim
10^{14}$. Suppose an electron ion pair is injected at radius
$R_o=x_oR_{Sch}$, where $R_{Sch}$ is the Schwarzchild radius. As
shown below, the electron proton pair is accelerated radially to
an asymptotic $\Gamma$ of $(l/x_o)^{1/3}$ when $\Gamma\gg1$. The
fractional power 1/3 is due to  the redshift suffered by the
radiation in the frame of the accelerated matter. Note that for
$l\sim 10^{14}$ and $x \sim 10^7$, the asymptotic $\Gamma$ is of
order 200, quite consistent with popular estimates.

At a collimating wall, however,  the baryonic outflow can be
constrained to be {\it non-radial}, and the radiation pressure can
be {\it more} effective because there is less redshift. Though
somewhat counterintuitive, this principle is similar to the fact
that a ship sails faster when kept oblique to the wind by its rudder
than when traveling directly downwind. If a collimating wall keeps
the flow moving oblique to the radial direction (which we are
assuming here is about the same as the direction of the photon wind)
then the acceleration or deceleration can be far more effective. The
extremely large super-Eddington factor l implies that the final
Lorentz factor is determined  by the collimation geometry.

The radiation force on an electron proton pair along its direction
of motion $\hat {\bf \beta}$, which moves at angle $\chi$ relative
to the radial direction $\hat r$, is most easily calculated in the
frame of the pair, where the cross section is the Thompson
cross-section. In this frame the component of force $\bf F^{\prime}$
along the direction of flow is given by
\begin{equation}
\bf F'\cdot{\bf \hat \beta} = \sigma_T {\bf \cal F'} \cdot {\bf\hat
\beta }/c \equiv \sigma_T\int d^2\hat k'
 \int_0^{\infty}I'(\nu', \hat k') (\hat k' \cdot  {\hat {\bf\beta}}) d\nu'/c
\end{equation}
where $ {\bf \cal F'}$  is the comoving radiative flux, $\sigma_T$
is the photon scattering cross section which throughout is assumed
to be the Thomson scattering cross section. The acceleration along
the wall in this frame, $a'$, is $a'=\bf{F' \cdot \hat \beta}/m$. In
the limit of strong radial combing of the radiation, $I'(\nu', \hat
k')$ can be factored into ${\bf \cal
F_{\nu}^{\prime}}(\nu')\delta^2(\hat k' - \hat r')$. Using the
invariance of $\int_0^{\infty}I'(\nu')d\nu'/\nu'^4$, we can write
\begin{equation}
{\cal F'_{\nu}}  = {\cal F_{\nu}}\frac{\delta^2(\hat k - \hat
r)}{\delta^2(\hat k' - \hat r')}\frac{\nu'^4}{\nu^4}
%(1+\beta)(1-\beta)^3\Gamma^4 = \frac{L}{4\pi
%r^2}(1+\beta)^{-2} \Gamma^{-2}
\end{equation}
where $\bf {\hat r'}$ is the k  vector of the radially combed
photons in the primed frame and  $\int_0^{\infty}{\cal F_{\nu}}d\nu
= l L_{edd}/4\pi r^2$. In the lab frame,
$d\beta\Gamma/dt=\Gamma^{3}a= a'$. The primed variables can be
expressed in terms of the unprimed variables as follows: Using the
aberration formula,
\begin{equation}
cos\chi'= [cos\chi - \beta]/[1- \beta cos\chi ]
\end{equation}
 and
\begin{equation}
\nu'=\nu \Gamma (1- \beta cos\chi),
\end{equation}
we can write
\begin{equation}
d\Gamma\beta/dt = \frac{lL_{Edd} \sigma_T}{4\pi r^2 m_p}
\frac{\delta^2(\hat k- \hat r)}{c \delta^2(\hat k' - \hat
r')}\Gamma^4(1-\beta cos\chi)^4 \frac {(cos\chi -\beta)} {(1-\beta
cos\chi)} .
\end{equation}

 When $\Gamma \gg 1$ and $1-cos\chi \ll 1-
\beta$, the quantity $cos \chi' =  (1-\beta cos\chi)^{-1}(cos\chi
-\beta) \sim 1 $, and $\delta^2(\hat k - \hat r)/
\delta^2(\hat k' - \hat r')\propto [d(1-\hat k' \cdot \hat r']/[d(1-\hat k
\cdot \hat r]
= (1+\beta)/(1- \beta) $, so
 \begin{equation}
d\Gamma\beta/dt \sim d\Gamma/dt = -(4 \Gamma)^{-2}  l d(1/x)
\label{acc}
\end{equation}
and the solution as a function of  radius $xR_{Sch}$ is
\begin{equation}
\Gamma(x) = [\frac{3l}{4} (1/x_o -1/x)+\Gamma_o^3]^{1/3}.
\end{equation}
This solution could be applicable to loose material that suddenly
finds itself in  the path of the fireball, such as a decaying
pickup neutron that has drifted in from the walls. For $x_o \sim
10^6$, corresponding to a radius of $\sim 10^{12}$cm, and $l\sim
10^{13}$ to $10^{14}$, this yields an asymptotic $\Gamma$ of
several hundred, which is an acceptable value, though Poynting
pressure could drive it to an even higher value. Also, the
Eddington luminosity increases with the number of pairs per
baryon, so l could effectively be higher than $10^{14}$.

 However,   for $l/x\gg \Gamma^3$, $\Gamma$ grows  very rapidly,
possibly  until the condition  $1-cos\chi \ll 1- \beta$ is no
longer satisfied. If the acceleration were to lead to   $1-cos\chi
\gg 1- \beta$, it  would follow that $1 -\beta cos\chi \sim 1-
cos\chi$, and the deceleration would then be  extremely powerful.
For $l/x\gg 1$, the value of $\Gamma$ is rapidly reduced to the
value where $cos\chi = \beta$, i.e. $\Gamma = 1/sin \chi$.

It is of course also possible that the condition $1-cos\chi \ll 1-
\beta$ is never satisfied, e.g. $\beta$ could be small and
$1-cos\chi$ could be of order unity. In this case the acceleration
along the wall is even larger than given by equation (\ref{acc})
because the radiation is not redshifted nearly as much; the
Doppler factor $\Gamma(1-\beta cos\chi)$ that would have lowered
the flux in the fluid frame when $1-cos\chi$ is negligible is now
less devastating. This only strengthens the conclusion that the
material is accelerated until (and only until) $\beta = cos\chi$.

The final value for $\Gamma$ is thus  $1/sin \chi$. By the same
argument, thermal motion is strongly  Compton cooled. If the
source has a finite size, then  the incident photons have some
finite angular spread in the  frame of the fluid.   However, any
Compton heating this may cause is modest, because the frequency in
the fluid frame $\nu'$ is only $\nu/\Gamma\ll m_ec^2$.  Moreover,
if the cylindrical radius of the fluid r is much larger than the
source size $r_s$, as we have assumed here, then the angular
spread of the photons in the fluid frame can be shown to be small,
of order $r_s/r$, (though larger than the angular extent of the
source as seen from the point of photon-fluid impact). A more
general treatment of finite source size will appear in a separate
paper  (Levinson and Eichler 2004).

 The incident radiation thus    makes
an angle of $\pi/2$ relative to the direction of motion, as would
wind relative to the motion of a frictionless sailboat whose
motion is kept by its rudder oblique to the wind direction. For
the extremely high super-Eddington factor l that is typical of
GRB's, we believe that viscous drag by the inner layers does not
change this conclusion significantly when the mean free scattering
time of an ion is longer than the acceleration time to $\beta =
cos \chi$.

The radiation, when scattered off the matter, is beamed most
strongly in the direction of motion, which corresponds to 90
degree scattering, and thus full polarization along the direction
$\hat n \times {\bf \beta}$. This should be contrasted with the
situation corresponding to the "head-on" approximation as termed
by Begelman and Sikora (1987) in which the radiation in the frame
of the scatterer is moving almost parallel to the lab frame
velocity as seen by the scatterer.  The head-on approximation
would be natural when the flow is ultra-relativistic and the angle
of incidence is physically unrelated to $\Gamma$. In this
circumstance, beaming along the direction of motion gives no
polarization, and it is only the finite spread of the beamed
radiation that allows polarized contributions  from scatterers
moving slightly off the line of sight.  For a hollow cone
geometry, the maximum intensity is seen by observers with lines of
sight on the cone. The  maximum polarization they may observe is
25 percent. The polarization vector is perpendicular to the cone,
i.e. each fluid element with direction $\hat \beta$ contributes
polarization in the direction $(\hat n-\hat\beta) \times \hat n$.
Sufficiently far away from the cone,  i.e. sufficiently far into
its interior or exterior, the polarization can be  larger than 25
percent, but only at the expense of lowered intensity. Thus the
$V_{max}$ associated with large polarizations is relatively small
and the probability of viewing polarization more than 75  percent,
say, is only 0.2 (Eichler and Levinson 2003).

By contrast, scattering from a Compton sail reverses the roles of
high and low polarization, so the largest $V_{max}$  is associated
with high polarization. In the limit of a line of sight of $\delta
\theta$ from the edge of an infinitely thin cone, where $\delta
\theta \ll \theta_0$, and $1/\Gamma \ll \theta_0$, most of the
contribution to the line of sight comes from fluid elements with
$\beta$ near $\hat n$, i.e. $\beta  - \hat n \sim \delta \theta$.

To set up a formal calculation, consider scattering off a thin
cone of opening angle $\theta_o$ defined by the azimuthal angle
$\phi$ running from 0 to $2\pi$, and the line of sight is offset
from the cone by angle $\Delta$. Let us use coordinates such that
the unit line of sight vector $ \hat n = [sin(\theta_0 + \Delta),
0, cos(\theta_0 + \Delta)]$  is in $x, z$ plane.  The ingoing
photon vector is $\vec k_1 = [sin(\theta_0 +\frac{1}{\Gamma})
cos\phi, sin(\theta_0 +\frac{1}{\Gamma})sin\phi, cos(\theta_0 +
\frac{1}{\Gamma})]$, the outgoing photon vector, $\vec k_2$ is parallel
to $\hat n$. The velocity vectors of the outflow, which form a
cone around the z axis, are defined by $\hat\beta = [sin\theta_0
cos\phi, sin\theta_0 sin\phi, cos\theta_0]$.

In the fully general 3-D scattering problem,  the degree of
polarization depends on the scattering angle $w'$, where $cos w' =
\hat k'_1 \cdot \hat k'_2$, whereas the Doppler boosting of the
intensity of the scattered radiation depends on  $\theta \equiv
arcos (\bf \hat \beta \cdot \bf \hat  n)$  as $I (\nu)=
I'(\nu')[\Gamma(1- \beta cos\theta)]^{-k}$. Here $k =3 -\alpha$
where $\alpha$ is the spectral index. The difficulty is that $w'$
and $\theta'$ are defined relative to different axes. However, in
the case that the incoming photon vector makes an angle of
$cos^{-1}\beta$ with the velocity vector, the frequency in the frame
of the scatterer is given by
\begin{equation}
\nu^{\prime} = \frac{\nu_1}{\Gamma},
\end{equation}
and we assume that $\nu'$  is conserved in the scattering.  We can use the
invariance of $k_{1\mu}k_2^{\mu}$ to write
\begin{equation}
1-cosw' = \Gamma^2 \frac{\nu_2}{\nu_1}(1-cosw) =
 \Gamma^2(1+\beta cos \theta^{\prime})(1-cosw).
\end{equation}

Using the aberration formula,
\begin{equation}
1+\beta cos\theta^{\prime} =  1+\beta \frac{(cos\theta-\beta)}{1 -
\beta cos\theta} = \frac{1}{\Gamma^2(1-\beta cos\theta)}
\end{equation}
we can write cos$w'$ in terms of laboratory coordinates as
\begin{equation}
cosw^{\prime} = 1-\frac{(1-cosw)}{(1-\beta cos\theta)}=
\frac{(cosw-\beta cos\theta)}{(1-\beta cos\theta)}.
\end{equation}

The Doppler boost and polarization expressions from emission from
a thin cone of opening angle $\theta_o$ can now be expressed as
functions of $\phi$, $\Gamma$ and $\theta_o$ using the basic
expressions

\begin{equation}
cosw=sin(\theta_0 +\Delta) sin(\theta_0 + \frac{1}{\Gamma})
cos\phi + cos(\theta_0 + \Delta)cos(\theta_0 + \frac{1}{\Gamma})
\end{equation}
%\\Outgoing photon $\vec K_2 = [sin(\theta_0 + \Delta), o, cos(\theta +
%\Delta)]$
and
\begin{equation}
\beta cos\theta = {\vec\beta} \cdot \hat n= \beta[sin\theta_0
sin(\theta_0 + \Delta)cos\phi+ cos\theta_0 cos(\theta_0 + \Delta)]
\end{equation}

In the limit of $1/\Gamma \ll \theta_o$, and $\Delta\ll \theta_o$,
we can expand the trigonometric formulae and obtain

\begin{equation}
1-cosw \sim[(1/\Gamma - \Delta)^2 +  \tilde\phi^2]/2,
\end{equation}
where $\tilde\phi= sin\theta_o\phi$, and \begin{equation} 1- \beta
cos\theta \sim (1/\Gamma^2 + \Delta^2 + \tilde\phi^2)/2
\end{equation}
whence
\begin{equation}
cosw-\beta cos\theta\sim \Delta/\Gamma
\end{equation}
Thus, in the viewing direction  $\Delta=0$, $cos w' =0$ and the
radiation scattered  into this direction after a single scattering
is nearly {\it fully polarized}. It can be further shown in the
single scattering limit that when $\Gamma \gg 1$, the polarization
vectors of all the rays scattered into a line of sight with
$\Delta = 0$ have polarization nearly perpendicular to the cone
axis as seen projected on the sky, so that their sum is nearly
fully polarized as well.

In the limit $1/\Gamma \ll \theta_o$ and $\Delta\ll \theta_o$, the
beaming direction  $\Delta =0$  corresponds to maximum
polarization. The direction of maximum total intensity (i.e. the
sum of intensities of both polarizations) does not coincide with
the beaming direction,  but even if we assume a detection
threshold for a gamma ray polarimeter that is a function of the
total intensity, the relative probability of detecting high
polarization is considerably higher than in the "head-on"
approximation. Writing the total intensity in the above limits,
where relevant $\phi$ are at $\phi \ll 1$, as
\begin{equation}
I(\Delta)\propto  \int_{-\pi}^{\pi} (1-\beta \cos
\theta)^{-k}(1+cosw'^2)sin\theta_o d\phi \sim
\int_{-\infty}^{\infty} (1-\beta \cos
\theta)^{-k}(1+cosw'^2)d\tilde\phi,
\end{equation}
we find, performing the integrals,  that
\begin{equation}
I(\Delta) \propto
[\Gamma^{-2}+\Delta^2]^{k-1/2}\left(1+\frac{4(2k+1)(2k-1)(\Delta\Gamma)^2}{(2k+2)(2k)[1+(\Delta
\Gamma)^2]^2 }\right)
\end{equation}
 For k=3,
\begin{equation}
\frac{I(\Delta= 0)}{I(\Delta= 1/\Gamma)}= 2^{5/2}\frac{48}{83}\sim
3.3,
\end{equation}
and, in a Euclidean space,
\begin{equation}
\frac{V_{max}(\Delta=0)}{V_{max}(\Delta=1/\Gamma)}\sim 6.
\end{equation}

 To summarize, scattering material in the
Compton sailing state beams the polarized scattered component
forward, and, for a roughly homogenous source distribution, a
nearly fully polarized beam has the highest $V_{max}$. The less
intense unpolarized component, which by invariance of emitted
power should be about as strong when averaged over a solid angle,
presumably covers a larger solid angle, so for a source
distribution with a low $\frac{V}{V_{max}}$, we might see  mostly
weakly-polarized GRB's. The above calculation has several
simplifying assumptions: it is for the analytically tractable case
$\theta_o \Gamma \gg1$, $\Delta\theta \ll \theta_o$ i.e.
scattering off a hollow cone into a thin annulus. It neglects
penetration to finite optical depths, multiple scattering, finite
source size, and viscosity within the sheath.  Many of these
issues will be addressed in a subsequent paper with detailed
numerical calculations (Levinson \& Eichler astro-ph/0402457).

Finally, it is worth recalling that the bulk Lorentz factor
$\Gamma$ of GRB fireballs, while usually estimated  from
observational constraints to be several hundred, has never been
calculated from first principles. The phenomenon  of Compton
sailing suggests that a) there may be a connection between
$\Gamma$ and the geometry of the flow and b) $\Gamma$ may vary
continuously between the fireball and the ejected,
supernova-associated material from the  envelope of the GRB's host
star. It suggests the possibility of calculating $\Gamma$
theoretically. While such a calculation would be formidable,  the
problem seems sufficiently well posed that $\Gamma$ is no longer a
mere free parameter imposed on the initial conditions of the
fireball.

I  thank A. Levinson,  E. Derishev and Y. Lyubarsky for useful
discussions. This research was supported by the Arnow Chair of
Astrophysics at Ben Gurion University, by an Adler Fellowship
awarded by the Israel Science Foundation, by a grant from the
Israel-U.S. Binational Science Foundation and by a Center of
Excellence grant from the Israeli Science Foundation.

\end{document}